\iftrue
\documentclass[aps,prl,twocolumn,
			   groupedaddress,superscriptaddress,
			   amsfonts,amssymb,amsmath,
			   citeautoscript,
			   a4paper]{revtex4-1}
\else
\documentclass[aps,pra,preprint,
			   groupedaddress,superscriptaddress,
			   amsfonts,amssymb,amsmath,
			   citeautoscript,
			   a4paper]{revtex4-1}
\fi

\usepackage{xcolor,soul}
\usepackage[pdftex]{hyperref}
\hypersetup{colorlinks,
			linkcolor={blue!75!black!80!yellow},
			citecolor={blue!75!black!80!yellow},
			urlcolor={blue!75!black!80!yellow},
			pdfstartview=FitH}
\usepackage{graphicx}
\usepackage{mathrsfs}
\usepackage{xspace}
\usepackage{braket}
\usepackage{xr}
\usepackage{gensymb} 
\usepackage{xcite}
\usepackage{stmaryrd} 
\usepackage[UKenglish]{babel}
\usepackage{placeins} 
\usepackage{bbm}
\usepackage{physics}
\usepackage{siunitx}
\usepackage{float}

\DeclareSIUnit[number-unit-product=]\percent{\char`\%} 

\usepackage{txfonts}  
\usepackage{txfontsb} 

\makeatletter
\newcommand*{\addFileDependency}[1]{
  \typeout{(#1)}
  \@addtofilelist{#1}
  \IfFileExists{#1}{}{\typeout{No file #1.}}
}
\makeatother

\makeatletter
\renewcommand\@make@capt@title[2]{%
	\@ifx@empty\float@link{\@firstofone}{\expandafter\href\expandafter{\float@link}}%
	\sffamily{\textbf{#1}}\@caption@fignum@sep#2
}%

\makeatother

\newcommand{\iu}{\mathrm{i}}

\newcommand{\appropto}{\mathrel{\vcenter{
			\offinterlineskip\halign{\hfil$##$\cr
				\propto\cr\noalign{\kern2pt}\sim\cr\noalign{\kern-2pt}}}}}

\usepackage{textcomp} 
\usepackage{xifthen}
\usepackage{etoolbox}
\newboolean{togglecomments}
\newboolean{togglechanges}

\setboolean{togglecomments}{true}
\setboolean{togglechanges}{false}

\usepackage[nameinlink, capitalize]{cleveref}
\newcommand{\hnamecref}[1]{\hyperref[#1]{\namecref{#1}}} 
\crefname{figure}{Figure}{Figures}
\Crefname{figure}{Fig.}{Figs.}
\Crefname{section}{Sec.}{Secs.}
\crefname{equation}{Eq.}{Eqs.}

\newcommand{\comment}[2]{%
    \ifbool{togglecomments}%
    {\textcolor{blue!70!black}{\small\textsf{%
    \textsuperscript{\textsc{\textsf{\MakeLowercase{#1}}}}%
    [#2]}}} 
    {}}     
\newcommand{\swap}[2]{\ifbool{togglechanges}
    {#2}  
    {\textcolor{red!70!black}{[#1]}\textrightarrow{}\textcolor{green!50!black}{[#2]}}}
\newcommand{\remove}[1]{\ifbool{togglechanges}
    {}    
    {\textcolor{red!70!black}{#1}}}
\newcommand{\inset}[1]{\ifbool{togglechanges}
    {#1}  
    {\textcolor{green!50!black}{#1}}}

\newcommand{\insetODM}[1]{\ifbool{togglechanges}
    {#1}  
    {\textcolor{red!50!green}{#1}}}
\newcommand{\insetIK}[1]{\ifbool{togglechanges}
    {#1}  
    {\textcolor{blue!40!white}{#1}}}
\newcommand{\insetZTX}[1]{\ifbool{togglechanges}
    {#1}  
    {\textcolor{cyan!60!green}{#1}}}

\newcommand{\optional}[1]{\ifbool{togglechanges}
    {}    
    {\textcolor{yellow!50!orange!80!gray}{#1}}}

\newcommand{\citeremind}[1]{%
    [\textcolor{blue!75!black!80!yellow}{
        $\blacksquare$%
	    \ifthenelse{\isempty{#1}}
	        {}
	        {\textsuperscript{\tiny\textsf{#1}}}%
	}]\xspace}

\newcommand{\hkuaffil}{\footnotesize Department of Physics and HK Institute of Quantum Science and Technology,\\The University of Hong Kong, Pokfulam, Hong Kong, China}
\newcommand{\natlabaffil}{\footnotesize State Key Laboratory of Optical Quantum Materials, The University of Hong Kong, Pokfulam, Hong Kong, China}

\begin{document}
\title{Observation of Floquet erratic non-Hermitian skin effect in photonic mesh lattice}

\author{Yeyang~Sun}
\thanks{Y.~S. and S.~Y. contributed equally to this work.}
\affiliation{\hkuaffil}
\affiliation{\natlabaffil}
\author{Shu~Yang}
\thanks{Y.~S. and S.~Y. contributed equally to this work.}
\affiliation{\hkuaffil}
\affiliation{\natlabaffil}
\author{Yi~Yang}
\email{yiyg@hku.hk}
\affiliation{\hkuaffil}
\affiliation{\natlabaffil}

\begin{abstract}
In ordered, translationally invariant non-Hermitian systems, the skin effect is understood as a boundary phenomenon: nonreciprocal hopping drives an extensive accumulation of eigenstates towards the edges, whereas the periodic-boundary spectrum remains Bloch extended.
Here we experimentally reveal the opposite limit---a disorder-enabled, boundary-independent, and intrinsically bulk form of skin localization---the recently predicted erratic non-Hermitian skin effect (ENHSE), realized in a driven photonic platform.
Using a time-multiplexed photonic mesh lattice with programmable gain, loss, and phase modulation, we engineer spatially fluctuating imaginary gauge fields and realize a Floquet non-Hermitian lattice whose global reciprocity can be tuned independently of strong local nonreciprocity. 
We observe a disorder-driven non-Hermitian topological transition between two oppositely directed disordered skin phases through a critical point of global reciprocity. At this transition, boundary skin accumulation disappears, yet the wave dynamics self-organizes into bulk-localized patterns without any interface, providing direct evidence of ENHSE. 
The measured localization profiles agree with simulations and exhibit the defining feature that distinct eigenstates share a common bulk-localized envelope determined by the disordered imaginary gauge fields. 
By further introducing controllable on-site disorder, we reveal the competition between ENHSE and Anderson localization, and show how increasing scattering progressively suppresses erratic skin dynamics. 
Our results help establish ENHSE as a unique disorder-induced non-Hermitian phenomenon and open a route to engineering localization, transport, and topology beyond conventional Bloch and boundary-based paradigms.
\end{abstract}

\maketitle

Non-Hermiticity offers a playground for open systems with gain, loss, and environmental coupling~\cite{el2018non,kawabata2019symmetry,bergholtz2021exceptional,ding2022non}.
It hosts a variety of unconventional phenomena with no direct Hermitian counterparts. 
Among them, the non-Hermitian skin effect (NHSE) has attracted particular attention~\cite{lee2016anomalous,yao2018edge,xiong2018does,kunst2018biorthogonal,alvarez2018non,yokomizo2019non,okuma2020topological,zhang2020correspondence,li2020critical,yi2020non,zhang2022review}. The NHSE is characterized by an extreme sensitivity of the spectrum and eigenstates to boundary conditions, leading to the extensive accumulation of eigenstates at the boundaries of a finite system. 
The emergence of skin modes is intimately tied to the spectral winding of complex energies~\cite{gong2018topological,zhang2020correspondence,okuma2020topological,wang2021generating}.
Moreover, this effect is closely related to the breakdown of conventional Hermitian bulk-boundary correspondence, spurring the development of non-Bloch band theory~\cite{yao2018edge,yao2018non,yokomizo2019non,yang2020non} that highlights the central role of translationally invariant band structures in the NHSE of ordered systems.

Going beyond this translationally invariant picture, the interplay of disorder and non-Hermiticity gives rise to new physical behavior~\cite{longhi2019topological,tang2020topological,liu2021real,weidemann2021coexistence,tzortzakakis2021transport,kim2021disorder,mo2022imaginary,weidemann2022topological,lin2022topological,leventis2022non,molignini2023anomalous,longhi2023anderson,guo2024scale,li2025universal}. 
This has led to intriguing discoveries such as the non-Hermitian Anderson insulator~\cite{tang2020topological,liu2021real,mo2022imaginary}, anomalous dynamics~\cite{weidemann2021coexistence,tzortzakakis2021transport,leventis2022non,longhi2023anderson,li2025universal}, and Lifshitz-tail states~\cite{silvestrov2001extended,longhi2025lifshitz}.
In particular, in non-Hermitian Anderson systems~\cite{weidemann2021coexistence,li2025universal}, the dynamics can exhibit jumping-like transitions between localized states, in sharp contrast to Hermitian Anderson localization, where wave-packet propagation is completely suppressed.
Focusing on the NHSE, although it is traditionally regarded as a consequence of translational invariance, growing attention has been devoted to its fate in disordered settings~\cite{longhi2019topological,kim2021disorder,claes2021skin,wanjura2021correspondence,lin2022topological,lin2022observation,dikopoltsev2022observation,sarkar2022interplay,liu2023modified,li2023disorder,zhang2023bulk,midya2024topological,jin2025anderson,wang2025observation}.
Non-Hermitian Aubry-Andre-Harper models have been shown to exhibit quasiperiodicity-driven skin-effect transitions~\cite{longhi2019topological,weidemann2022topological}, whereas generic random disorder typically weakens or suppresses NHSE~\cite{weidemann2020topological,lin2022observation}.
By contrast, suitably tailored disorder can induce skin accumulation, underscoring the nontrivial role of spatial inhomogeneity in non-Hermitian systems~\cite{claes2021skin,kim2021disorder,sarkar2022interplay,li2025universal,wang2025observation}.

Experimental investigation of such non-Hermitian localization phenomena requires a platform with flexible control over coupling, gain/loss, and disorder.
Synthetic photonic mesh lattices~\cite{regensburger2011photon,regensburger2012parity,wimmer2015observation,xiao2017observation,chen2018observation,chalabi2019synthetic,weidemann2020topological,weidemann2021coexistence,wimmer2021superfluidity,weidemann2022topological,steinfurth2022observation,marques2023observation,yu2024dirac,qin2024temporal,monika2025quantum,feis2025space,pang2025topological,wang2025nonlinear} provide an ideal platform in this regard.
In these systems, the evolution of light is governed by discrete-time equations, leading to an effective Floquet Hamiltonian.
This time-multiplexed platform has attracted extensive attention, enabling the realization of parity-time-symmetric synthetic photonic lattices~\cite{regensburger2012parity}, the observation of parity-time-symmetric solitons~\cite{wimmer2015observation} and topological edge states~\cite{xiao2017observation,chen2018observation,feis2025space}, as well as quantum walks in synthetic gauge fields~\cite{chalabi2019synthetic,pang2025topological}. 
Owing to their high tunability, photonic mesh lattices have been widely used to study a variety of Hermitian and non-Hermitian disordered phenomena, including topological phase transitions~\cite{weidemann2022topological}, unusual forms of non-Hermitian transport~\cite{weidemann2021coexistence}, and induced transparency~\cite{steinfurth2022observation}. 

Very recently, a new type of disorder-induced localized state in non-Hermitian systems has been proposed, and was termed the erratic non-Hermitian skin effect (ENHSE)~\cite{longhi2025erratic}.
ENHSE is induced by spatially fluctuating imaginary gauge fields~\cite{claes2021skin,midya2024topological,longhi2025erratic,nan2026anomalous} and features eigenstates that display a consistent spatial profile across the bulk, while the associated Lyapunov exponent vanishes, indicating the absence of exponential localization, which distinguishes ENHSE from both conventional NHSE and Anderson localization. 
Despite these intriguing features, the experimental realization of this recently predicted ENHSE remains elusive.

\begin{figure}[htbp]
    \centering
    \includegraphics[width=\linewidth]{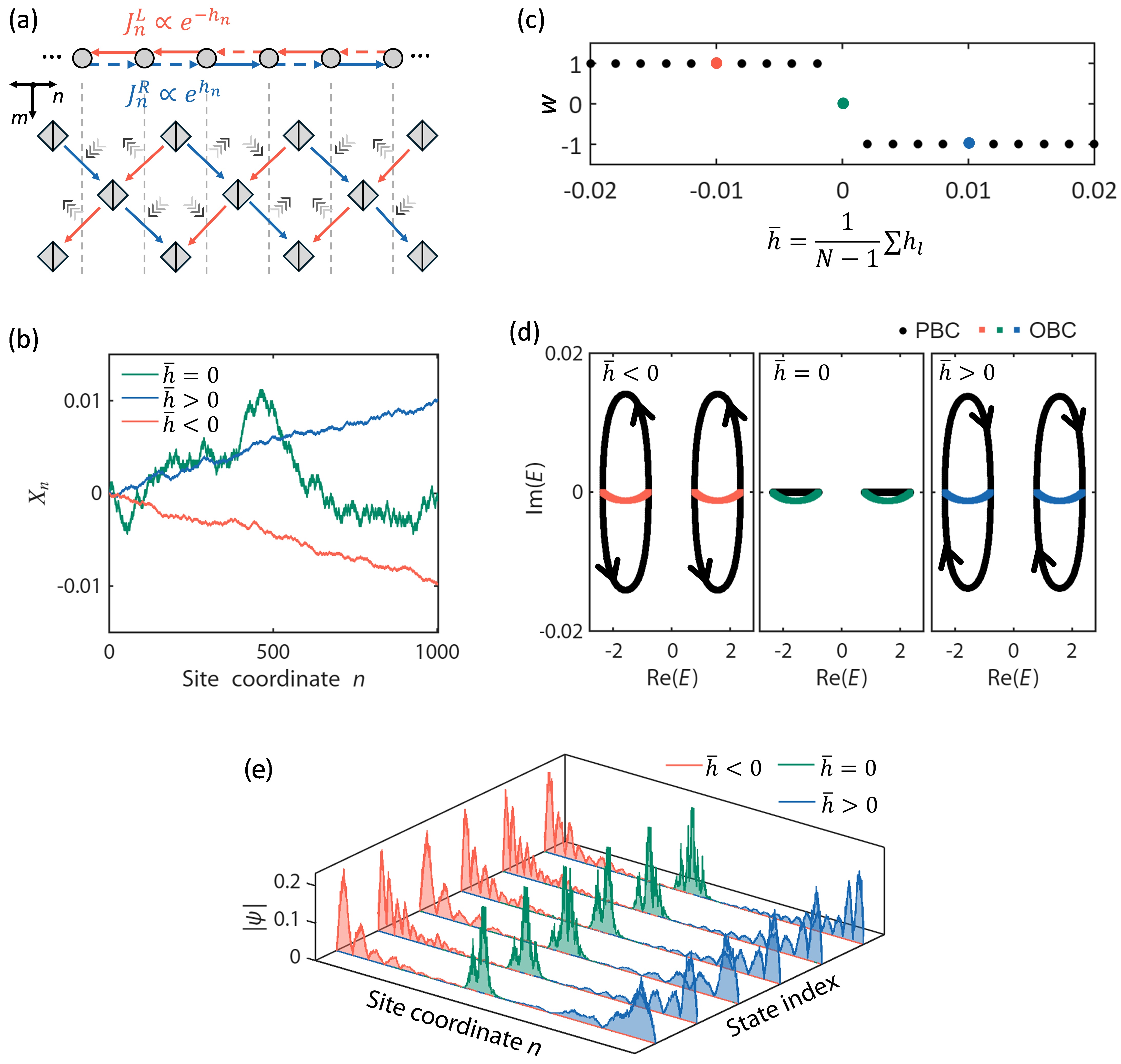}
    \caption{
    \textbf{Erratic non-Hermitian skin effect as a non-Hermitian phase transition featuring bulk localization.}
    \textbf{a.~}Schematic of a one-dimensional Hatano-Nelson lattice (top) with nonreciprocal left- and right-hopping amplitudes $J_n^{L}\propto\exp(-h_n)$ and $J_n^{R}\propto\exp(h_n)$ that can be described by disordered imaginary gauge fields $\{h_n\}$.
    A Floquet realization (bottom), based on a time-multiplexed quantum walk in a photonic mesh lattice, encodes the disorder in the time-varying gain and loss (shaded gray arrows) in the conditioned translation (red and blue arrows)
    \textbf{b.~}The normalized sum of disordered imaginary gauge fields $X_n=\frac{1}{N-1}\sum_{l=1}^{n-1}{h_l}$ ($X_1\equiv0$).
    \textbf{c.~}Real-space winding number $W$ versus the average imaginary gauge fields $\bar{h}$. A non-Hermitian topological phase transition occurs around $\bar{h}=0$, where $W$ switches from 1 to -1 through $W=0$. 
    \textbf{d.~}Complex energy spectra under PBC (black) and OBC (colored) for three cases: $\bar{h}<0$, $\bar{h}=0$, and $\bar{h}>0$. 
    \textbf{e.~}Spatial distributions of localized OBC eigenstates for different $\bar{h}$ corresponding to d. For $\bar{h}<0$ and $\bar{h}>0$, the states are localized near opposite boundaries, corresponding to winding numbers $w=\pm1$. For $\bar{h}=0$, all the OBC eigenstates localize around the same position in the bulk.
    }
    \label{fig:1}
\end{figure}

In this work, we report an experimental observation of ENHSE in the Floquet dynamics of a photonic mesh lattice.
We realize disordered imaginary gauge fields---a crucial ingredient for ENHSE---through programming the time-dependent amplified or attenuated coupling coefficients between two coupled fiber loops. 
By ingeniously tailoring the driven sequences of multiple amplitude modulators, we engineer the gauge-field fluctuations and demonstrate ENHSE as a non-Hermitian topological phase transition in the disordered non-Hermitian skin effect (DNHSE). 
Finally, we show how dynamics can be shaped by the simultaneous presence and competition between Anderson localization and ENHSE.

\begin{figure*}[htbp]
    \centering
    \includegraphics[width=0.65\linewidth]{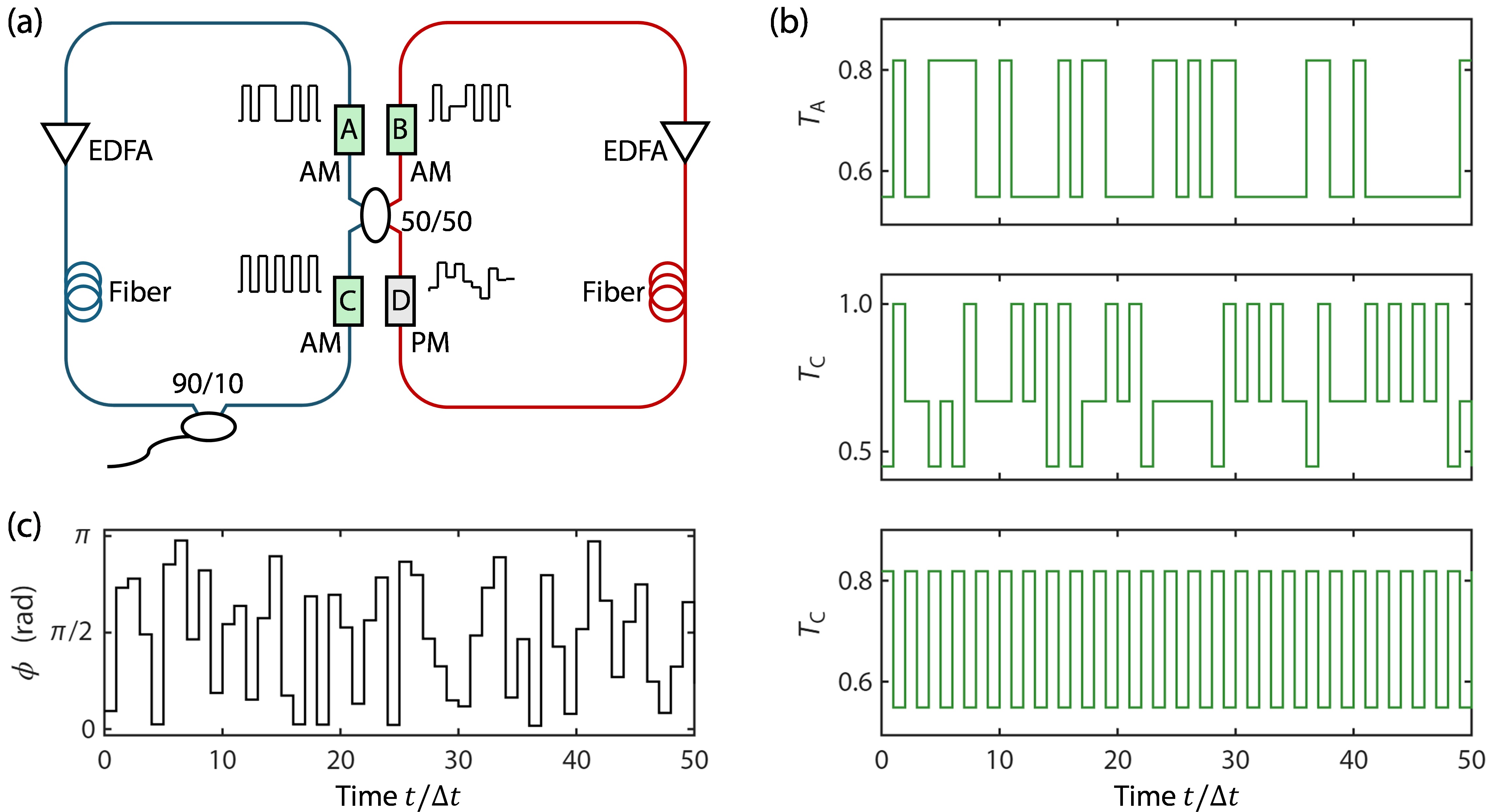}
    \caption{
    \textbf{Experimental setup and disordered modulation waveforms.}
    \textbf{a.~}A photonic mesh lattice setup showing two fiber loops of different lengths coupled by a beam splitter.  Both loops contain erbium-doped fiber amplifiers (EDFAs) to compensate for the insertion loss. The disordered imaginary gauge fields are introduced via the three amplitude modulators (AMs). The phase modulator (PM) allows for the control of real on-site potentials. 
    \textbf{b.~}Example transmittance modulation sequence of the three AMs.
    \textbf{c.~}Example phase modulation of the PM.  $\Delta t$ is half of the round-trip time difference of pulse propagation inside the two loops.
    }
    \label{fig:2}
\end{figure*}

We first illustrate the origin of ENHSE with the disordered Hatano-Nelson model~\cite{hatano1996localization,longhi2025erratic} [shown in \Cref{fig:1}(a) top], described by the non-Hermitian tight-binding Hamiltonian $\hat{H}=\sum_{n=1}^{N-1}\left(J_n^R \hat{c}_{n+1}^{\dagger}\hat{c}_n+ J_n^L \hat{c}_{n}^{\dagger}\hat{c}_{n+1}\right)$, where $\hat{c}_n^\dagger$ is the particle creation operator at site $n$, and $N$ is the number of lattice sites.
Disorder manifests in the form of imaginary gauge fields~\cite{midya2024topological,longhi2025erratic}---the imbalanced left- and right-hopping amplitudes are $J_n^L=J \exp(-h_n)$ and $J_n^R=J \exp(h_n)$, where $J$ is a constant and $h_n$ obeys a distribution $\{h_n\}$ of statistical mean $\bar{h}$.
$\bar{h}$ quantifies the global reciprocity, which describes the net directional bias across the entire lattice. The system can maintain global reciprocity while simultaneously being nonreciprocal locally.
The non-Hermitian localization properties are determined by the global reciprocity $\bar{h}$ [see Supplemental Material (SM)] rather than the specific disorder distribution.
In the main text, we focus on $\{h_n\}$ that obeys the Bernoulli distribution (see additional results under the uniform and normal distributions in the SM). 

To experimentally demonstrate ENHSE, we consider the platform of photonic mesh lattice [shown in \Cref{fig:1}(a) bottom]. 
This Floquet platform is primarily composed of two fiber loops with unequal lengths, which are connected through a 50:50 fiber coupler.
The pulse dynamics in the two loops can be described by a one-dimensional time-bin--encoded discrete lattice. 
Based on this platform, our objective is therefore to construct a time-evolution operator that corresponds to a non-Hermitian Hamiltonian with disordered imaginary gauge fields that exhibits global non-reciprocity or reciprocity.
The basic evolution is given by two coupled equations:
\begin{subequations}\label{eq:evolution}
\begin{align}
        u^{m+1}_n & = \left(G_{uu}u^m_{n+1}+\mathrm{i}G_{uv}v^m_{n+1}\right)e^{\mathrm{i}\phi_n}, \\
        v^{m+1}_n & =\mathrm{i}G_{vu}u^m_{n-1}+G_{vv}v^m_{n-1}\,,
\end{align}
\end{subequations}
where $G_{\alpha\beta}(n)\left(\alpha,\beta\in\left\{u,v\right\}\right)$ are the four spatially-disordered attenuated or amplified coupling coefficients among left movers (labeled by $u$) and right movers (labeled by $v$) at site $n$, $\phi_n$ is a tunable phase factor at site $n$, and $u^m_n$ and $v^m_n$ represent the pulse amplitudes at lattice position $n$ and time bin $m$ in the left and right loops, respectively. 

We derive the explicit form of the time-evolution operator $U$ by applying an inverse similarity transformation $U=VU_0V^{-1}$ on a uniform quantum-walk system $U_0=SR(\beta)$, where $S=\sum_n\ket{n-1}\bra{n}\otimes\ket{u}\bra{u}+\ket{n}\bra{n-1}\otimes\ket{v}\bra{v}$ is the conditional translation operator, $R(\beta) = \exp(\iu\beta\sigma_x)$ is the coin operator, $\beta=\pi/4$ for equal probability of coin toss, and $V=\mathrm{diag}\left[1,\exp(\sum_{l=1}^{1} h_l),\exp(\sum_{l=1}^{2} h_l),\cdots,\exp(\sum_{l=1}^{N-1} h_l)\right]$
is the local gauge transformation with the phase factor determined by the random sequence $\{h_n\}$.
Based on this transformation, we obtain the expressions for attenuation and amplification coefficients:
\begin{subequations}\label{eq:4G}
\begin{align}
        G_{uu}(n)&=\cos\left(\beta \right)\exp\left(-\sum\nolimits_{2n}^{2n+1}h_l\right)\,,\\
        G_{uv}(n)&=\sin\left(\beta \right)\exp\left(-\sum\nolimits_{2n}^{2n+2}h_l\right)\,,\\
        G_{vu}(n)&=\sin\left(\beta \right)\exp\left(\sum\nolimits_{2n-2}^{2n}h_l\right)\,,\\
        G_{vv}(n)&=\cos\left(\beta \right)\exp\left(\sum\nolimits_{2n-1}^{2n}h_l\right)\,.
\end{align}
\end{subequations}
Notably, the gain and loss arrangement at a particular site $n$ is determined by the open path integral of the disordered link variables in its vicinity.

\cref{fig:1}(b) presents the normalized integrated imaginary gauge fields---$X_1\equiv0$ and $X_n=\frac{1}{N-1}\sum_{l=1}^{n-1} h_l$ for $n>1$---of three sets of random imaginary gauge fields with the mean exponent $\bar{h}$ greater than, less than, and equal to zero, respectively.
As the summation adds up, the blue (red) partial sum profile exhibits an overall increasing (decreasing) trend for $\bar{h}>0$ ($\bar{h}<0$) in spite of small fluctuations, indicating the accumulation of wave functions at the right (left) boundary. 
In contrast, the green partial sum curve fluctuates around $\bar{h}=0$ (green), suggesting the absence of boundary localization. 

\begin{figure*}[htbp]
    \centering
    \includegraphics[width=1\textwidth]{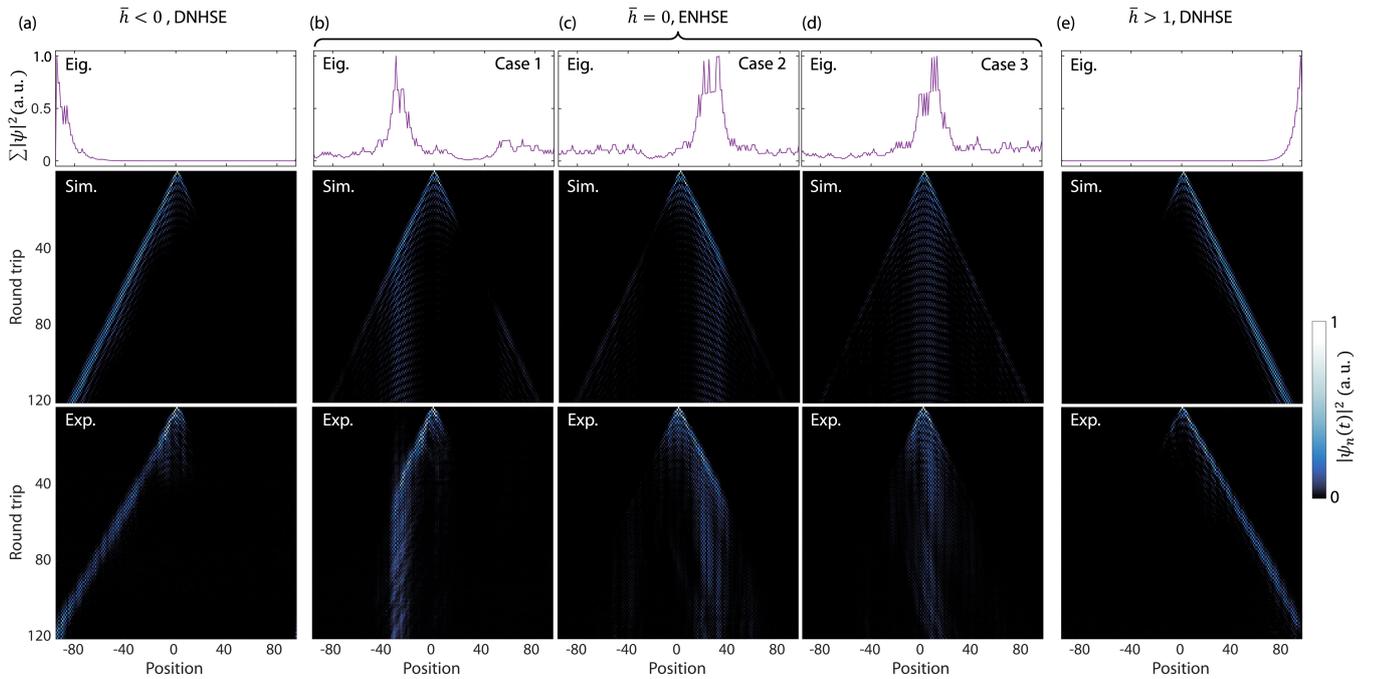}
    \caption{
    \textbf{Floquet dynamics of erratic non-Hermitian skin effect.}
    \textbf{a.~}Disordered non-Hermitian skin effect (DNHSE) under global nonreciprocity $\bar{h}<0$.    
    \textbf{b-d.~}Three examples of erratic non-Hermitian skin effect under global reciprocity $\bar{h}=0$, which exhibit a defining feature of bulk localization without the need for an interface.    
    \textbf{e.~}Disordered non-Hermitian skin effect under global nonreciprocity $\bar{h}>0$.
    Top: Summed modulus-squared eigenstate distribution; middle: simulated dynamics; bottom: measured dynamics.
}
    \label{fig:3}
\end{figure*}

In general, the boundary localization induced by the NHSE can be diagnosed by the spectral winding number~\cite{kawabata2019symmetry,okuma2020topological,zhang2020correspondence,bergholtz2021exceptional}.
For such disordered lattices, it is more convenient to perform the real-space calculation of the winding number~\cite{song2019non,claes2021skin} (see SM):
\begin{equation}
    w(E_0)=\cfrac{1}{N} \mathrm{Tr}(\hat{Q}^{\dagger}[\hat{Q},\hat{X}]) 
\end{equation}
where $E_0$ is the reference complex energy, $\hat{X}$ is the position operator, and the unitary operator $\hat{Q}$ is defined by the polar decomposition $\hat{H}-E_0=\hat{Q}\hat{P}$, with $\hat{P}$ being a positive semi-definite Hermitian matrix.
\cref{fig:1}(c) shows the winding number of our Floquet system \cref{eq:evolution} as a function of disorder $\bar{h}$. 
The winding numbers are quantized to $+1$ and $-1$ for $\bar{h}<0$ and $\bar{h}>0$, respectively, indicating the appearance of non-Hermitian skin localization even with disordered imaginary gauge fields. 
Of particular interest is their phase transition at $\bar{h}=0$, which corresponds to a zero winding number; this phase transition and the associated wave localization phenomena are the primary goals of our experimental effort.

To corroborate the real-space winding number calculation, \Cref{fig:1}(d) provides the Floquet energy spectra for the three cases under periodic boundary conditions (PBCs) and open boundary conditions (OBCs). 
When $\bar{h}<0$ or $\bar{h}>0$, within each of the two Floquet bands, the PBC spectra wind around the OBC arc counterclockwise (winding number +1) and clockwise (winding number -1), respectively. 
At the transition $\bar{h}=0$, the PBC spectra themselves collapse into open arcs, similar to their OBC counterparts.

\cref{fig:1}(e) illustrates examples of the eigenstate spatial distribution.
The skin localizations on either edge of the open chain are marked with green ($\bar{h}<0$) and blue ($\bar{h}>0$), respectively.
For $\bar{h}=0$, the vanishing winding number in \Cref{fig:1}(d) indicates the absence of the conventional NHSE. Indeed, all eigenstates cease to accumulate at the boundary; instead, they form a distinct class of bulk-localized states, recently theoretically predicted as the ENHSE~\cite{longhi2025erratic}.
The center of the localized wavefunction can occur anywhere in the bulk without requiring an interface.
Moreover, the profile of the bulk-localized wavefunctions is highly dependent on the given random sequences $\{h_n\}$, and, more precisely, the resulting integrated imaginary gauge field $X_n$.
Unlike Anderson localization, where different eigenstates generally have common exponential localizations yet distinct localization centers, all ENHSE eigenstates share the same spatial distribution and non-exponential localization. 

Compared with previous studies of photonic mesh lattices~\cite{regensburger2011photon,weidemann2020topological,weidemann2021coexistence,steinfurth2022observation,weidemann2022topological}, here the major experimental challenge is to achieve independent control of the gain and loss across all the four possible coupling channels, namely, $G_{uu}(n)$, $G_{uv}(n)$, $G_{vu}(n)$, $G_{vv}(n)$, between the two loops.
These are four amplitude operations, but we can always normalize them by one of the $G_{u\left(v\right)u\left(v\right)}\left(n\right)$ in the linear experimental regime. 
Therefore, in the experimental setup, we add three amplitude modulators (AMs), labeled by A, B, and C in \Cref{fig:2}(a), driven by different time-varying signals inside the two loops.
The round-trip time of the optical wave packet in the long and short loop is $\bar{t}+\Delta t$ and $\bar{t}-\Delta t$, respectively, where $\bar{t}\approx\SI{24.7}{\micro\second} $ and $\Delta t \approx \SI{0.13}{\micro\second}$.
\cref{fig:2}(b) shows the beginning part of a programmed transmittance $T_{A,B,C}$ example of the three AMs for the target $\{h_n\}$ [green line in \Cref{fig:1} (b)] satisfying the Bernoulli distribution of zero mean (see SM). 
Meanwhile, a phase modulator (PM), labeled by $D$ in \Cref{fig:2}(a), is also added to introduce the tunability of the real potential energy to the one-dimensional lattice via the phase delay $\phi_n$.  
\cref{fig:2}(c) shows an example time-varying phase delay waveform, which provides a control knob for the disorder in the on-site potential.
Taken together, the incorporation of three AMs and one PM enables a comprehensive control over the hopping strength disorder [gray gradient arrows in \Cref{fig:1}(a)] and on-site potential disorder in the mesh lattice.

We first demonstrate pristine ENHSEs in the absence of on-site disorder by setting $\phi_n=0$.
\cref{fig:3} presents the eigenstates together with the simulated and measured dynamics for five examples: one for $\bar{h}<0$, one for $\bar{h}>0$, and three for $\bar{h}=0$, all of which adopt an initial condition of single-site excitation. 
When $\bar{h}<0$ [\Cref{fig:3}(a)], the injected light pulse keeps propagating to the left, showing similar evolution characteristics to the conventional NHSE in the absence of hopping disorder. 
As $\bar{h}$ increases and reaches the balance point of global reciprocity $\bar{h}=0$ [\Cref{fig:3}(b-d)], the ENHSE emerges. 
After a certain number of round-trips, the incident wave packet forms a stable localization pattern inside the bulk without the presence of an interface, yet the detailed shape of the localization depends on the particulars of the disordered hopping that maintains global reciprocity [\Cref{fig:3}(b-d) and see SM for more realizations]. 
Across \Cref{fig:3}(b-d), the observed localization patterns (bottom) of the three examples all agree well with their associated eigenstate (top) and simulated dynamics (middle).
As we further increase $\bar{h}>0$ [\Cref{fig:3}(e)], we observe the restoration of NHSE of opposite winding number to that of \Cref{fig:3}(a), and, thus, all the eigenstates are now propagating rightward.
We emphasize that these ENHSE localized states are distinct from those in NHSE and non-Hermitian Anderson localization. 
On the one hand, NHSE requires the existence of boundaries or interfaces within the system to generate localized states, whereas ENHSE's localization can occur at any possible location within the bulk. 
On the other hand, non-Hermitian Anderson localization features localization switching (see SM in the temporal dynamics because of the excitation of spatially distinct localized modes, but this switching is absent in ENHSE because of the spatial resemblance of all bulk localized eigenstates.

\begin{figure*}[htbp]
    \centering
    \includegraphics[width=\linewidth]{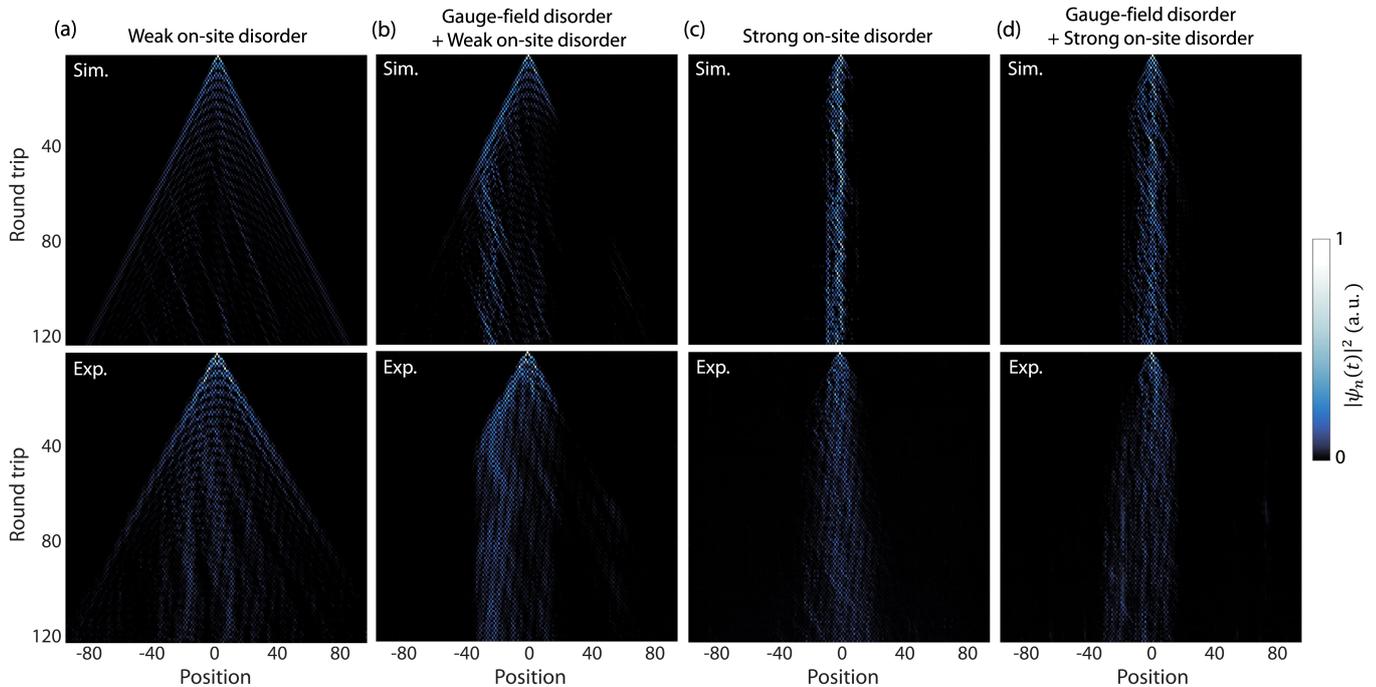}
    \caption{
    \textbf{Floquet dynamics shaped by disordered imaginary gauge fields and disordered on-site potential.}
    \textbf{a.~}Ballistic spreading of wavefunction under weak on-site disorder with $\phi_{\mathrm{max}}=0.1\pi$.
    \textbf{b.~}Evident ENHSE in the presence of the weak on-site disorder (a).
    \textbf{c.~}Anderson localization under strong on-site disorder with $\phi_{\mathrm{max}}=\pi$.
    \textbf{d.~}Suppressed ENHSE in the presence of the strong on-site disorder (c).
    The AM modulations in (b) and (d) are configured in the same way as those in \Cref{fig:3}(b).
    Top: simulated dynamics; bottom: measured dynamics.
    \label{fig:4}
    }
\end{figure*}

Next, we explore the difference and interplay between ENHSE and Anderson localization by turning on the on-site disorder $\phi_n$.
Based on the PM, we employ a uniformly distributed phase disorder $\phi_n \in [0,\phi_{\mathrm{max}}]$ with a tunable maximal phase delay $\phi_{\mathrm{max}}$ dictated by the voltage applied to the PM.
Under weak disorder ($\phi_{\mathrm{max}}=0.1\pi$) and in the absence of imaginary gauge fields, the wave packet can still spread over a considerable distance, exhibiting extended-like dynamics as shown in \Cref{fig:4}(a).
Therefore, such weak disorder still allows for the occurrence of the ENHSE dynamical process, and the erratic localization remains evident despite the simultaneous presence of wavefunction fluctuations and additional spreading, as shown in \Cref{fig:4}(b).
Nevertheless, as the on-site disorder is further increased to $\phi_{\mathrm{max}}=\pi$ [\Cref{fig:4}(c)], the wave-packet dynamics becomes fully localized, resulting from multiple scattering induced by the on-site disorder. 
Such suppression of transport substantially blurs the ENHSE-induced localization, causing the dynamics to become increasingly dominated by Anderson localization instead. [\Cref{fig:4}(d)].
Based on the comparison of the four scenarios, we shall see the competition between disordered imaginary gauge fields and on-site disorder: stronger on-site disorder results in enhanced scattering, obstructs the integration of the disordered link variables for manifesting the effect of global reciprocity, and, therefore, drives the system into an Anderson-localized regime.

In summary, we have experimentally realized the ENHSE in a time-multiplexed photonic mesh lattice with programmable disordered imaginary gauge fields. By engineering spatially fluctuating gain and loss across four coupling channels, we demonstrated a disorder-driven non-Hermitian topological transition---diagnosed by a real-space winding number---between two oppositely directed skin phases through a globally reciprocal critical point at which all eigenstates collapse onto a common bulk-localized profile without requiring any boundary or interface. 
This behavior, together with the agreement between measured dynamics, numerical simulations, and eigenstate profiles, constitutes experimental evidence for ENHSE in a synthetic Floquet setting.
Our results help establish ENHSE as an experimentally accessible, intrinsically bulk, and disorder-enabled non-Hermitian phenomenon that falls outside both Bloch-band and boundary-based frameworks, and suggest future directions, including the exploration of ENHSE in higher dimensions, its interplay with nonlinearity, and applications in disorder-engineered wave control.

\textit{Acknowledgement:} We thank Xiang Ji, Qingyang Mo, Li Zhang for useful discussion, and Tianxiang Dai, Lingrui Hong, Yitian Tong, Yu Xia, and Chao Xiang for experimental help.

\textit{Note added:} During the completion of the project, we became aware of related work on the acoustic platform~\cite{zhong2026observation}.

\bibliographystyle{apsrev4-2}
\bibliography{references}

@PREAMBLE{
 "\providecommand{\noopsort}[1]{}" 
 # "\providecommand{\singleletter}[1]{#1}%" 
}

@article{bergholtz2021exceptional,
  title={Exceptional topology of non-Hermitian systems},
  author={Bergholtz, Emil J and Budich, Jan Carl and Kunst, Flore K},
  journal={Reviews of Modern Physics},
  volume={93},
  number={1},
  pages={015005},
  year={2021},
  publisher={APS},
}

@article{yi2020non,
  title={Non-Hermitian Skin Modes Induced by On-Site Dissipations and Chiral Tunneling Effect},
  author={Yi, Yifei and Yang, Zhesen},
  journal={Physical Review Letters},
  volume={125},
  number={18},
  pages={186802},
  year={2020},
  publisher={APS}
}

@article{regensburger2012parity,
  title={Parity--time synthetic photonic lattices},
  author={Regensburger, Alois and Bersch, Christoph and Miri, Mohammad-Ali and Onishchukov, Georgy and Christodoulides, Demetrios N and Peschel, Ulf},
  journal={Nature},
  volume={488},
  number={7410},
  pages={167--171},
  year={2012},
  publisher={Nature Publishing Group UK London},
}

@article{kawabata2019symmetry,
  title={Symmetry and topology in non-Hermitian physics},
  author={Kawabata, Kohei and Shiozaki, Ken and Ueda, Masahito and Sato, Masatoshi},
  journal={Physical Review X},
  volume={9},
  number={4},
  pages={041015},
  year={2019},
  publisher={APS},
}

@article{lee2016anomalous,
  title={Anomalous edge state in a non-Hermitian lattice},
  author={Lee, Tony E},
  journal={Physical Review Letters},
  volume={116},
  number={13},
  pages={133903},
  year={2016},
  publisher={APS},
}

@article{zhong2026observation,
  title={Observation of Erratic Non-Hermitian Skin Localization and Transport},
  author={Zhong, Jia-Xin and Kim, Jee Woo and Longhi, Stefano and Jing, Yun},
  journal={arXiv preprint arXiv:2601.19749},
  year={2026}
}

@article{nan2026anomalous,
  title={Anomalous Localization and Mobility Edges in Non-Hermitian Quasicrystals with Disordered Imaginary Gauge Fields},
  author={Nan, Guolin and Li, Zhijian and Mei, Feng and Xu, Zhihao},
  journal={arXiv preprint arXiv:2601.14754},
  year={2026}
}

@article{chen2018observation,
  title={Observation of topologically protected edge states in a photonic two-dimensional quantum walk},
  author={Chen, Chao and Ding, Xing and Qin, Jian and He, Yu and Luo, Yi-Han and Chen, Ming-Cheng and Liu, Chang and Wang, Xi-Lin and Zhang, Wei-Jun and Li, Hao and others},
  journal={Physical Review Letters},
  volume={121},
  number={10},
  pages={100502},
  year={2018},
  publisher={APS}
}

@article{chalabi2019synthetic,
  title={Synthetic gauge field for two-dimensional time-multiplexed quantum random walks},
  author={Chalabi, Hamidreza and Barik, Sabyasachi and Mittal, Sunil and Murphy, Thomas E and Hafezi, Mohammad and Waks, Edo},
  journal={Physical Review Letters},
  volume={123},
  number={15},
  pages={150503},
  year={2019},
  publisher={APS}
}

@article{xiao2017observation,
  title={Observation of topological edge states in parity--time-symmetric quantum walks},
  author={Xiao, L and Zhan, X and Bian, ZH and Wang, KK and Zhang, X and Wang, XP and Li, J and Mochizuki, K and Kim, D and Kawakami, N and others},
  journal={Nature Physics},
  volume={13},
  number={11},
  pages={1117--1123},
  year={2017},
  publisher={Nature Publishing Group UK London}
}

@article{wimmer2015observation,
  title={Observation of optical solitons in PT-symmetric lattices},
  author={Wimmer, Martin and Regensburger, Alois and Miri, Mohammad-Ali and Bersch, Christoph and Christodoulides, Demetrios N and Peschel, Ulf},
  journal={Nature Communications},
  volume={6},
  number={1},
  pages={7782},
  year={2015},
  publisher={Nature Publishing Group UK London}
}

@article{li2020critical,
  title={Critical non-Hermitian skin effect},
  author={Li, Linhu and Lee, Ching Hua and Mu, Sen and Gong, Jiangbin},
  journal={Nature Communications},
  volume={11},
  number={1},
  pages={5491},
  year={2020},
  publisher={Nature Publishing Group UK London},
}

@article{yao2018non,
  title={Non-hermitian chern bands},
  author={Yao, Shunyu and Song, Fei and Wang, Zhong},
  journal={Physical Review Letters},
  volume={121},
  number={13},
  pages={136802},
  year={2018},
  publisher={APS},
}

@article{yokomizo2019non,
  title={Non-Bloch band theory of non-Hermitian systems},
  author={Yokomizo, Kazuki and Murakami, Shuichi},
  journal={Physical Review Letters},
  volume={123},
  number={6},
  pages={066404},
  year={2019},
  publisher={APS},
}

@article{song2019non,
  title={Non-Hermitian topological invariants in real space},
  author={Song, Fei and Yao, Shunyu and Wang, Zhong},
  journal={Physical Review Letters},
  volume={123},
  number={24},
  pages={246801},
  year={2019},
  publisher={APS},
}

@article{zhang2020correspondence,
  title={Correspondence between winding numbers and skin modes in non-Hermitian systems},
  author={Zhang, Kai and Yang, Zhesen and Fang, Chen},
  journal={Physical Review Letters},
  volume={125},
  number={12},
  pages={126402},
  year={2020},
  publisher={APS},
}

@article{dikopoltsev2022observation,
  title={Observation of Anderson localization beyond the spectrum of the disorder},
  author={Dikopoltsev, Alex and Weidemann, Sebastian and Kremer, Mark and Steinfurth, Andrea and Sheinfux, Hanan Herzig and Szameit, Alexander and Segev, Mordechai},
  journal={Science Advances},
  volume={8},
  number={21},
  pages={eabn7769},
  year={2022},
  publisher={American Association for the Advancement of Science}
}

@article{regensburger2011photon,
  title={Photon propagation in a discrete fiber network: An interplay of coherence and losses},
  author={Regensburger, Alois and Bersch, Christoph and Hinrichs, Benjamin and Onishchukov, Georgy and Schreiber, Andreas and Silberhorn, Christine and Peschel, Ulf},
  journal={Physical Review Letters},
  volume={107},
  number={23},
  pages={233902},
  year={2011},
  publisher={APS},
}

@article{tzortzakakis2021transport,
  title={Transport and spectral features in non-Hermitian open systems},
  author={Tzortzakakis, AF and Makris, KG and Szameit, A and Economou, EN},
  journal={Physical Review Research},
  volume={3},
  number={1},
  pages={013208},
  year={2021},
  publisher={APS}
}

@article{mo2022imaginary,
  title={Imaginary-disorder-induced topological phase transitions},
  author={Mo, Qingyang and Sun, Yeyang and Li, Junkai and Ruan, Zhichao and Yang, Zhaoju},
  journal={Physical Review Applied},
  volume={18},
  number={6},
  pages={064079},
  year={2022},
  publisher={APS}
}

@article{liu2021real,
  title={Real-space topological invariant and higher-order topological Anderson insulator in two-dimensional non-Hermitian systems},
  author={Liu, Hongfang and Zhou, Ji-Kun and Wu, Bing-Lan and Zhang, Zhi-Qiang and Jiang, Hua},
  journal={Physical Review B},
  volume={103},
  number={22},
  pages={224203},
  year={2021},
  publisher={APS}
}

@article{tang2020topological,
  title={Topological Anderson insulators in two-dimensional non-Hermitian disordered systems},
  author={Tang, Ling-Zhi and Zhang, Ling-Feng and Zhang, Guo-Qing and Zhang, Dan-Wei},
  journal={Physical Review A},
  volume={101},
  number={6},
  pages={063612},
  year={2020},
  publisher={APS}
}

@article{li2025universal,
  title={Universal non-Hermitian transport in disordered systems},
  author={Li, Bo and Chen, Chuan and Wang, Zhong},
  journal={Physical Review Letters},
  volume={135},
  number={3},
  pages={033802},
  year={2025},
  publisher={APS}
}

@article{longhi2025erratic,
  title={Erratic non-Hermitian skin localization},
  author={Longhi, Stefano},
  journal={Physical Review Letters},
  volume={134},
  number={19},
  pages={196302},
  year={2025},
  publisher={APS}
}

@article{steinfurth2022observation,
  title={Observation of photonic constant-intensity waves and induced transparency in tailored non-Hermitian lattices},
  author={Steinfurth, Andrea and Kre{\v{s}}i{\'c}, Ivor and Weidemann, Sebastian and Kremer, Mark and Makris, Konstantinos G and Heinrich, Matthias and Rotter, Stefan and Szameit, Alexander},
  journal={Science Advances},
  volume={8},
  number={21},
  pages={eabl7412},
  year={2022},
  publisher={American Association for the Advancement of Science}
}

@article{pang2025topological,
  title={Topological quantum walk in synthetic non-Abelian gauge fields with photonic mesh lattices},
  author={Pang, Zehai and Abdelghani, Omar and Solja{\v{c}}i{\'c}, Marin and Yang, Yi},
  journal={Optica},
  volume={12},
  number={11},
  pages={1794--1799},
  year={2025},
  publisher={Optica Publishing Group}
}

@article{midya2024topological,
  title={Topological phase transition in fluctuating imaginary gauge fields},
  author={Midya, Bikashkali},
  journal={Physical Review A},
  volume={109},
  number={6},
  pages={L061502},
  year={2024},
  publisher={APS}
}

@article{zhang2022review,
  title={A review on non-Hermitian skin effect},
  author={Zhang, Xiujuan and Zhang, Tian and Lu, Ming-Hui and Chen, Yan-Feng},
  journal={Advances in Physics: X},
  volume={7},
  number={1},
  pages={2109431},
  year={2022},
  publisher={Taylor \& Francis},
}

@article{okuma2020topological,
  title={Topological origin of non-Hermitian skin effects},
  author={Okuma, Nobuyuki and Kawabata, Kohei and Shiozaki, Ken and Sato, Masatoshi},
  journal={Physical Review Letters},
  volume={124},
  number={8},
  pages={086801},
  year={2020},
  publisher={APS},
}

@article{gong2018topological,
  title={Topological phases of non-Hermitian systems},
  author={Gong, Zongping and Ashida, Yuto and Kawabata, Kohei and Takasan, Kazuaki and Higashikawa, Sho and Ueda, Masahito},
  journal={Physical Review X},
  volume={8},
  number={3},
  pages={031079},
  year={2018},
  publisher={APS},
}

@article{yao2018edge,
  title={Edge states and topological invariants of non-Hermitian systems},
  author={Yao, Shunyu and Wang, Zhong},
  journal={Physical Review Letters},
  volume={121},
  number={8},
  pages={086803},
  year={2018},
  publisher={APS},
}

@article{yang2020non,
  title={Non-Hermitian bulk-boundary correspondence and auxiliary generalized Brillouin zone theory},
  author={Yang, Zhesen and Zhang, Kai and Fang, Chen and Hu, Jiangping},
  journal={Physical Review Letters},
  volume={125},
  number={22},
  pages={226402},
  year={2020},
  publisher={APS}
  
}

@article{kunst2018biorthogonal,
  title={Biorthogonal bulk-boundary correspondence in non-Hermitian systems},
  author={Kunst, Flore K and Edvardsson, Elisabet and Budich, Jan Carl and Bergholtz, Emil J},
  journal={Physical Review Letters},
  volume={121},
  number={2},
  pages={026808},
  year={2018},
  publisher={APS},
}

@article{alvarez2018non,
  title={Non-Hermitian robust edge states in one dimension: Anomalous localization and eigenspace condensation at exceptional points},
  author={Alvarez, VM Martinez and Vargas, JE Barrios and Torres, LEF Foa},
  journal={Physical Review B},
  volume={97},
  number={12},
  pages={121401},
  year={2018},
  publisher={APS},
}

@article{kim2021disorder,
  title={Disorder-driven phase transition in the second-order non-Hermitian skin effect},
  author={Kim, Kyoung-Min and Park, Moon Jip},
  journal={Physical Review B},
  volume={104},
  number={12},
  pages={L121101},
  year={2021},
  publisher={APS}
}

@article{zhang2023bulk,
  title={Bulk-boundary correspondence in disordered non-Hermitian systems},
  author={Zhang, Zhi-Qiang and Liu, Hongfang and Liu, Haiwen and Jiang, Hua and Xie, XC},
  journal={Science Bulletin},
  volume={68},
  number={2},
  pages={157--164},
  year={2023},
  publisher={Elsevier}
}

@article{liu2023modified,
  title={Modified generalized Brillouin zone theory with on-site disorder},
  author={Liu, Hongfang and Lu, Ming and Zhang, Zhi-Qiang and Jiang, Hua},
  journal={Physical Review B},
  volume={107},
  number={14},
  pages={144204},
  year={2023},
  publisher={APS}
}

@article{li2023disorder,
  title={Disorder-induced entanglement phase transitions in non-Hermitian systems with skin effects},
  author={Li, Kai and Liu, Ze-Chuan and Xu, Yong},
  journal={arXiv preprint arXiv:2305.12342},
  year={2023}
}

@article{claes2021skin,
  title={Skin effect and winding number in disordered non-Hermitian systems},
  author={Claes, Jahan and Hughes, Taylor L},
  journal={Physical Review B},
  volume={103},
  number={14},
  pages={L140201},
  year={2021},
  publisher={APS}
}

@article{wang2025observation,
  title={Observation of disorder-induced boundary localization},
  author={Wang, Bing-Bing and Cheng, Zheyu and Zou, Hong-Yu and Ge, Yong and Zhao, Ke-Qi and Si, Qiao-Rui and Yuan, Shou-Qi and Sun, Hong-Xiang and Xue, Haoran and Zhang, Baile},
  journal={Proceedings of the National Academy of Sciences},
  volume={122},
  number={19},
  pages={e2422154122},
  year={2025},
  publisher={National Academy of Sciences}
}

@article{jin2025anderson,
  title={Anderson delocalization in strongly coupled disordered non-Hermitian chains},
  author={Jin, Wei-Wu and Liu, Jin and Wang, Xin and Zhang, Yu-Ran and Huang, Xueqin and Wei, Xiaomin and Ju, Wenbo and Yang, Zhongmin and Liu, Tao and Nori, Franco},
  journal={Physical Review Letters},
  volume={135},
  number={7},
  pages={076602},
  year={2025},
  publisher={APS}
}

@article{sarkar2022interplay,
  title={Interplay of disorder and point-gap topology: Chiral modes, localization, and non-Hermitian Anderson skin effect in one dimension},
  author={Sarkar, Ronika and Hegde, Suraj S and Narayan, Awadhesh},
  journal={Physical Review B},
  volume={106},
  number={1},
  pages={014207},
  year={2022},
  publisher={APS}
}

@article{wanjura2021correspondence,
  title={Correspondence between non-Hermitian topology and directional amplification in the presence of disorder},
  author={Wanjura, Clara C and Brunelli, Matteo and Nunnenkamp, Andreas},
  journal={Physical Review Letters},
  volume={127},
  number={21},
  pages={213601},
  year={2021},
  publisher={APS}
}

@article{molignini2023anomalous,
  title={Anomalous skin effects in disordered systems with a single non-Hermitian impurity},
  author={Molignini, Paolo and Arandes, Oscar and Bergholtz, Emil J},
  journal={Physical Review Research},
  volume={5},
  number={3},
  pages={033058},
  year={2023},
  publisher={APS}
}

@article{guo2024scale,
  title={Scale-tailored localization and its observation in non-Hermitian electrical circuits},
  author={Guo, Cui-Xian and Su, Luhong and Wang, Yongliang and Li, Li and Wang, Jinzhe and Ruan, Xinhui and Du, Yanjing and Zheng, Dongning and Chen, Shu and Hu, Haiping},
  journal={Nature Communications},
  volume={15},
  number={1},
  pages={9120},
  year={2024},
  publisher={Nature Publishing Group UK London}
}

@article{qin2024temporal,
  title={Temporal Goos-H{\"a}nchen shift in synthetic discrete-time heterolattices},
  author={Qin, Chengzhi and Wang, Shulin and Wang, Bing and Hu, Xinyuan and Liu, Chenyu and Li, Yinglan and Zhao, Lange and Ye, Han and Longhi, Stefano and Lu, Peixiang},
  journal={Physical Review Letters},
  volume={133},
  number={8},
  pages={083802},
  year={2024},
  publisher={APS}
}

@article{el2018non,
  title={Non-Hermitian physics and PT symmetry},
  author={El-Ganainy, Ramy and Makris, Konstantinos G and Khajavikhan, Mercedeh and Musslimani, Ziad H and Rotter, Stefan and Christodoulides, Demetrios N},
  journal={Nature Physics},
  volume={14},
  number={1},
  pages={11--19},
  year={2018},
  publisher={Nature Publishing Group UK London}
}

@article{wang2025nonlinear,
  title={Nonlinear non-Hermitian skin effect and skin solitons in temporal photonic feedforward lattices},
  author={Wang, Shulin and Wang, Bing and Liu, Chenyu and Qin, Chengzhi and Zhao, Lange and Liu, Weiwei and Longhi, Stefano and Lu, Peixiang},
  journal={Physical Review Letters},
  volume={134},
  number={24},
  pages={243805},
  year={2025},
  publisher={APS}
}

@article{wimmer2021superfluidity,
  title={Superfluidity of light and its breakdown in optical mesh lattices},
  author={Wimmer, Martin and Monika, Monika and Carusotto, Iacopo and Peschel, Ulf and Price, Hannah M},
  journal={Physical Review Letters},
  volume={127},
  number={16},
  pages={163901},
  year={2021},
  publisher={APS}
}

@article{feis2025space,
  title={Space-time-topological events in photonic quantum walks},
  author={Feis, Joshua and Weidemann, Sebastian and Sheppard, Tom and Price, Hannah M and Szameit, Alexander},
  journal={Nature Photonics},
  volume={19},
  number={5},
  pages={518--525},
  year={2025},
  publisher={Nature Publishing Group UK London}
}

@article{monika2025quantum,
  title={Quantum state processing through controllable synthetic temporal photonic lattices},
  author={Monika, Monika and Nosrati, Farzam and George, Agnes and Sciara, Stefania and Fazili, Riza and Marques Muniz, Andr{\'e} Luiz and Bisianov, Arstan and Lo Franco, Rosario and Munro, William J and Chemnitz, Mario and others},
  journal={Nature Photonics},
  volume={19},
  number={1},
  pages={95--100},
  year={2025},
  publisher={Nature Publishing Group UK London}
}

@article{yu2024dirac,
  title={Dirac mass induced by optical gain and loss},
  author={Yu, Letian and Xue, Haoran and Guo, Ruixiang and Chan, Eng Aik and Terh, Yun Yong and Soci, Cesare and Zhang, Baile and Chong, YD},
  journal={Nature},
  volume={632},
  number={8023},
  pages={63--68},
  year={2024},
  publisher={Nature Publishing Group UK London}
}

@article{marques2023observation,
  title={Observation of photon-photon thermodynamic processes under negative optical temperature conditions},
  author={Marques Muniz, AL and Wu, FO and Jung, PS and Khajavikhan, M and Christodoulides, DN and Peschel, U},
  journal={Science},
  volume={379},
  number={6636},
  pages={1019--1023},
  year={2023},
  publisher={American Association for the Advancement of Science}
}

@article{weidemann2022topological,
  title={Topological triple phase transition in non-Hermitian Floquet quasicrystals},
  author={Weidemann, Sebastian and Kremer, Mark and Longhi, Stefano and Szameit, Alexander},
  journal={Nature},
  volume={601},
  number={7893},
  pages={354--359},
  year={2022},
  publisher={Nature Publishing Group UK London}
}

@article{lin2022topological,
  title={Topological phase transitions and mobility edges in non-Hermitian quasicrystals},
  author={Lin, Quan and Li, Tianyu and Xiao, Lei and Wang, Kunkun and Yi, Wei and Xue, Peng},
  journal={Physical Review Letters},
  volume={129},
  number={11},
  pages={113601},
  year={2022},
  publisher={APS}
}

@article{xiong2018does,
  title={Why does bulk boundary correspondence fail in some non-hermitian topological models},
  author={Xiong, Ye},
  journal={Journal of Physics Communications},
  volume={2},
  number={3},
  pages={035043},
  year={2018},
  publisher={IOP Publishing},
}

@article{wang2021generating,
  title={Generating arbitrary topological windings of a non-Hermitian band},
  author={Wang, Kai and Dutt, Avik and Yang, Ki Youl and Wojcik, Casey C and Vu{\v{c}}kovi{\'c}, Jelena and Fan, Shanhui},
  journal={Science},
  volume={371},
  number={6535},
  pages={1240--1245},
  year={2021},
  publisher={American Association for the Advancement of Science},
}

@article{hatano1996localization,
  title={Localization transitions in non-Hermitian quantum mechanics},
  author={Hatano, Naomichi and Nelson, David R},
  journal={Physical Review Letters},
  volume={77},
  number={3},
  pages={570},
  year={1996},
  publisher={APS},
}

@article{lin2022observation,
  title={Observation of non-Hermitian topological Anderson insulator in quantum dynamics},
  author={Lin, Quan and Li, Tianyu and Xiao, Lei and Wang, Kunkun and Yi, Wei and Xue, Peng},
  journal={Nature Communications},
  volume={13},
  number={1},
  pages={3229},
  year={2022},
  publisher={Nature Publishing Group UK London}
}

@article{weidemann2020topological,
  title={Topological funneling of light},
  author={Weidemann, Sebastian and Kremer, Mark and Helbig, Tobias and Hofmann, Tobias and Stegmaier, Alexander and Greiter, Martin and Thomale, Ronny and Szameit, Alexander},
  journal={Science},
  volume={368},
  number={6488},
  pages={311--314},
  year={2020},
  publisher={American Association for the Advancement of Science},
}

@article{weidemann2021coexistence,
  title={Coexistence of dynamical delocalization and spectral localization through stochastic dissipation},
  author={Weidemann, Sebastian and Kremer, Mark and Longhi, Stefano and Szameit, Alexander},
  journal={Nature Photonics},
  volume={15},
  number={8},
  pages={576--581},
  year={2021},
  publisher={Nature Publishing Group UK London}
}

@article{longhi2025lifshitz,
  title={Lifshitz tail states in non-Hermitian disordered photonic lattices},
  author={Longhi, Stefano},
  journal={Optics Letters},
  volume={50},
  number={3},
  pages={746--749},
  year={2025},
  publisher={Optica Publishing Group}
}

@article{silvestrov2001extended,
  title={Extended tail states in an imaginary random potential},
  author={Silvestrov, PG},
  journal={Physical Review B},
  volume={64},
  number={7},
  pages={075114},
  year={2001},
  publisher={APS}
}

@article{longhi2023anderson,
  title={Anderson localization in dissipative lattices},
  author={Longhi, Stefano},
  journal={Annalen der Physik},
  volume={535},
  number={5},
  pages={2200658},
  year={2023},
  publisher={Wiley Online Library}
}

@article{leventis2022non,
  title={Non-Hermitian jumps in disordered lattices},
  author={Leventis, A and Makris, KG and Economou, EN},
  journal={Physical Review B},
  volume={106},
  number={6},
  pages={064205},
  year={2022},
  publisher={APS}
}

@article{ding2022non,
  title={Non-Hermitian topology and exceptional-point geometries},
  author={Ding, Kun and Fang, Chen and Ma, Guancong},
  journal={Nature Reviews Physics},
  volume={4},
  number={12},
  pages={745--760},
  year={2022},
  publisher={Nature Publishing Group UK London},
}

@article{longhi2019topological,
  title={Topological phase transition in non-Hermitian quasicrystals},
  author={Longhi, Stefano},
  journal={Physical Review Letters},
  volume={122},
  number={23},
  pages={237601},
  year={2019},
  publisher={APS}
}

\end{document}